\documentclass[final, 5p]{elsarticle}
\usepackage{amsmath,amssymb}
\usepackage{lineno}
\newcommand\Tstrut{\rule{0pt}{2.6ex}}  
\usepackage{enumitem}
\usepackage{textcomp}
\modulolinenumbers[5]
\usepackage{supertabular, booktabs}
\usepackage{threeparttable}
\usepackage{dcolumn}
\usepackage{epstopdf}
\usepackage[figuresright]{rotating}
\usepackage{xcolor}
\usepackage{multirow}
\usepackage{tabularx}
\usepackage{mathtools}

\newcolumntype{s}{>{\centering\arraybackslash}X}

\newcommand{\at}[2][]{#1|_{#2}}

\journal{Acta Materialia}
\biboptions{sort&compress}

\begin{document}

\begin{frontmatter}

\title{Quantum mechanics basis of quality control in hard metals}

\author[ampkth]{Ruiwen Xie \corref{cor1}}
\cortext[cor1]{Corresponding author}
\ead{ruiwen2@kth.se}

\author[ampkth]{Raquel Liz\'arraga \corref{cor2}}
\cortext[cor2]{Corresponding author}
\ead{raqli@kth.se}
\author[ampkth]{David Linder}
\author[ampkth]{Ziyong Hou}
\author[ampkth]{Valter Str\"om}
\author[sandvik]{Martina Lattemann}
\author[sandvik]{Erik Holmstr\"om}
\author[ampkth]{Wei Li}

\author[ampkth,pauu,ispohu]{Levente Vitos}

\address[ampkth]{Applied Materials Physics, Department of Materials Science and Engineering, Royal Institute of Technology, Stockholm SE-10044, Sweden}
\address[sandvik]{Sandvik Coromant R\&D, Stockholm SE-12680, Sweden.}

\address[pauu]{Department of Physics and Astronomy, Division of Materials Theory, Uppsala University, Uppsala, Sweden}
\address[ispohu]{Wigner Research Centre for Physics, Institute for Solid State Physics and Optics, Budapest, Hungary}

\begin{abstract}
Non-destructive and reliable quality control methods are a key aspect to designing, developing and
manufacturing new materials for industrial applications and new technologies.
The measurement of the magnetic saturation is one of such methods and it is conventionally 
employed in the cemented carbides industry. 
We present a general quantum mechanics based relation between the magnetic saturation and 
the components of the binder phase of cemented carbides, 
which can be directly employed as a quality control. To illustrate our results, 
we calculate the magnetic saturation of a binder phase, 85Ni15Fe binary alloy, using ab-initio methods 
and compare the theoretical predictions to the magnetic saturation measurements. 
We also analyse interface and segregation effects on the magnetic saturation 
by studying the electronic structure of the binder phase. The excellent agreement between calculations 
and measurements demonstrates the applicability of our method to any binder phase. Since the magnetic saturation is employed
to ensure the quality of cemented carbides, the present method allows us to explore new materials for alternative binder phases efficiently.
\end{abstract}

\begin{keyword}
\texttt 
Hard metal \sep binder phase \sep magnetic saturation \sep ab-initio calculations
\end{keyword}

\end{frontmatter}


\section{Introduction}

Cemented carbides, also known as hard metals, are relatively tough and fatigue resistant 
materials used in cutting tools and rock drilling inserts, among other industrial applications. 
Essentially, hard metals are composite materials made of tungsten carbide (WC) grains glued together by a binder phase. Typically, the binder phase consists of ductile cobalt and some amount of dissolved tungsten and carbon. 
The microstructure of these materials is predominantly controlled by the components of the binder phase. 
For example, the grain growth rate is slower in W-rich binders, while faster and more homogeneous 
in C-rich binders~\cite{chabretou1999quantitative}. Moreover, the morphology and polymorphism of the WC-Co alloys are shown to be significantly influenced by the dissolved amount of W in the binder phase~\cite{marshall2015role, weidow2010binder}. Therefore, the knowledge of the composition of the binder phase, can be used to predict microstructural properties of hard metals.

Despite the great success of Co as a binder phase, the need of finding
a substitute has been recognized because of cobalt's rising price and 
health threats~\cite{jobs1940powder,linna2004exposure}. 
One important criterion for an appropriate substitute is that the 
range of the total C concentration in the sample, the so-called carbon window, should be relatively broad 
for the sake of industrial manufacturing processes, so that the undesired $\eta$-phase (M$ _{x} $C) and graphite can be avoided easily.
A series of WC-(Fe,Co,Ni) cemented carbides, which possesses comparable properties to Co-bonded hard metals, 
such as hardness, wear resistance and strength, was developed by Prakash \textit{et al.}~\cite{prakash1977properties,prakash1979properties}. In addition, the Fe-Ni-W-C phase diagram calculated by Guillermet demonstrated that the carbon window widens with the increase of Ni concentration~\cite{guillermet1987assessment}. 
The WC-(Fe, Mn) hard metals were also reported to be promising alternative binders due to their higher hardness and comparative fracture toughness~\cite{hanyaloglu2001production}.  
Recently, a cutting insert made of a new cemented carbide with 
a high entropy alloy as a binder phase showed superior performance in a 
machining test compared to a state-of-the-art cutting insert made of a 
Co-bonded hard metal~\cite{holmstrom2018high,lizarraga2018}.

The measurement of the magnetic saturation is commonly used to identify the composition of the binder phase and to determine the localization of the sample in the carbon window, which is an important indicator of the properties of the composite~\cite{brookes1998hardmetals,love2010quantitative}. 
By measuring the change of the magnetic saturation, the W concentration dissolved into the binder phase and the overall C content in the composite can be non-destructively estimated~\cite{love2006hard}.
Therefore, this measurement is routinely employed in the cutting tool industry as a quality control for cemented carbides.
Such a quantitative relationship between the magnetic saturation and W composition in the binder
has been constructed for WC-Co alloys~\cite{roebuck1984influence,love2010quantitative}. 
Unfortunately, today there is no such relation for a more general binder consisting of an alloy, which 
severely hampers the development of more complex binder phases.  
Hence, to obtain such a relationship is critical for the production of new Co-free cemented carbides.   

In the present study, we formulate and compute a quantitative relationship between the magnetic saturation and the components of the binder phase using ab-initio methods. As an example, we use a binary 85Ni15Fe alloy as a binder to illustrate this relationship, 
however, our modelling method is general and can be applied to more complex multi-component alloys. 
The theoretical results are then compared to experimental measurements of the weight specific magnetic saturation to demonstrate the applicability of our model.
We also investigate the effects of metal/ceramic interfaces and segregation of the binder elements to 
the dissolved components in the binder phase on the magnetic saturation. 
 
The paper is organized as follows: in Sec.~\ref{sec:method} we derive the mathematical model
to calculate the weight specific magnetic saturation
and in Sec.~\ref{CompModel}, we introduce the
computational model. Ab-initio methods are described 
in Sec.~\ref{sec:definitions}.
Results and discussions are shown in Sec.~\ref{sec:Results} and finally in Sec.~\ref{Conclu} we summarize our findings.    

\section{\label{sec:method}Theory}

The carbon level of cemented carbides can be determined in a non-destructive way by measuring the weight specific magnetic
saturation, $\sigma_s$, as discussed in Ref.~\cite{love2006hard}. Early work by Roebuck and Almond on Co-W-C alloys, 
demonstrated that the magnetic saturation decreases linearly with respect to W content~\cite{roebuck1984influence}. Here we derive an expression
to calculate the weight specific magnetic saturation for cemented carbides
with a generic binder and we calculate it using ab-initio methods.

The weight specific magnetic 
saturation can be expressed as

\begin{equation}
\sigma_s= \frac{\sum_i \mu_i}{\sum_i m_i},
\end{equation}
where $\mu_i$ and $m_i$ are the atomic magnetic moments and the masses of all the elements in the material, respectively. In
particular, a hard metal only contains magnetic specimens in the binder, 
and if we consider an ideal binder, in which neither C- nor W-diffusion occurs into the binder phase,
the magnetic saturation can be written as

\begin{equation}
\sigma^b_s = \frac{\sum_j^{N^b} \mu^b_j}{\sum_i m_i} = \frac{N^b \mu^b}{\sum_i m_i}.
\label{sigma}
\end{equation}
The index $j=\{1,N^b\}$, where $N^b$ is the number of atoms in an ideal binder,
the quantity $\mu_j^b$ is the atomic magnetic moment of a binder atom without diffusion of W(C) and 
$\mu^b$ is the average magnetic moment of the binder atoms.
However, both W and C do dissolve into the binder during sintering, causing the magnetic moments 
of their neighbouring binder atoms to change. 
The effect of the dissolved W and C on the magnetic saturation is then described by the ratio,

\begin{equation}
  \begin{aligned}
 \frac{\sigma_s}{\sigma^b_s}= \frac{\sum_i \mu_i}{\sum_j \mu^b_j}
&=\frac{N^b \mu^b + N_W \mu^{\text{eff}}_W + N_c \mu^{\text{eff}}_C}{N^b \mu^b}\\
&=1+\frac{N_W \mu^{\text{eff}}_W}{N^b \mu^b} + \frac{N_c \mu^{\text{eff}}_C}{N^b \mu^b}.
\label{COM}
 \end{aligned}
\end{equation}
The quantities $N_W$ and $N_C$ are the number of W and C atoms 
dissolved in the binder, respectively.
The variables $\mu^{\text{eff}}_W$ and $\mu^{\text{eff}}_C$ correspond to the effective
magnetic moment changes induced by W and C, respectively and we define them as 

\begin{subequations}
\begin{equation}
 \mu^{\text{eff}}_W =  \mu_W + \delta \mu_W
\label{delta-W}
\end{equation}
and
\begin{equation}
 \mu^{\text{eff}}_C =  \mu_C + \delta \mu_C,
\label{delta-C}
 \end{equation}
\end{subequations}
where  $\delta \mu_{W} $ and $ \delta \mu_{C} $ denote
the W-induced and C-induced magnetic moment changes of the binder host, respectively. 
$\mu_W$ and $\mu_C$ are the magnetic moments of W and C in the binder phase, respectively.
One can re-write Eq.~\ref{COM} using the atomic concentration of W and C ($C_W$ and $C_C$) 
in the binder to obtain the following expression

\begin{equation}
  \begin{aligned}
 \frac{\sigma_s}{\sigma^b_s} =1  &  + \left (\frac{C_W}{1-C_W-C_C}\right) 
\frac{\mu^{\text{eff}}_W}{ \mu^b}  \\  &  \qquad + \left (\frac{C_C}{1-C_W-C_C}\right) 
\frac{\mu^{\text{eff}}_C}{ \mu^b}.
 \end{aligned}
\label{COM1}
\end{equation}
%
By ignoring the contribution of C 
to the weight specific magnetic saturation, $C_{C} = 0$ in Eq.~\ref{COM1} 
and assuming a small concentration of W in the binder, one recovers the
linear phenomenological expression derived 
by Roebuck and Almond for Co binders~\cite{roebuck1984influence,Lattermann2018}.

In the following section we present the computational models to simulate the binder phase and calculate $\mu^{\text{eff}}_{W}$ and $\mu^{\text{eff}}_{C} $ from first principles.

\subsection{Computational model}
\label{CompModel}

\begin{figure}[t]
	\begin{center}
		\begin{tabular}{c@{\qquad}c}
			\resizebox{\columnwidth}{!}{\includegraphics[clip]{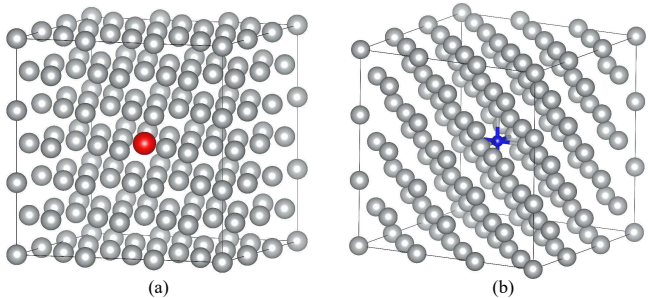}}
			\\
		\end{tabular}
		\caption{\label{fig:NiWC} (Colour online) The computational models (a) W-substitutional binder phase and (b) C-interstitial binder phase, in which the red and blue spheres correspond to 
W and C atoms, respectively. The grey spheres are binder atoms. 
The number of binder atoms, $N_b$, is 107 in a) and 108 in b).}
	\end{center}
\end{figure}

In order to calculate $\mu^{\text{eff}}_{W} $ and $\mu^{\text{eff}}_{C}$ in Eqs.~\ref{delta-W}
and ~\ref{delta-C} we use the computational models shown in Fig.~\ref{fig:NiWC}a) and b), respectively.
In our calculations we adopt the face-centred cubic (fcc) structure as a model, however our method is applicable to any
crystal structure. 
Figure~\ref{fig:NiWC}a) displays a 3$\times$3$\times$3 fcc supercell 
for a binder, 85Ni15Fe, in which one binder atom has been substituted
by a W atom, while Fig.~\ref{fig:NiWC}b) corresponds to a supercell where one C atom occupies
the octahedral interstitial position, which is verified to be the most stable occupation. 
In the following we assume that
the dissolved W and C atoms do not interact with each other in the binder phase.
This assumption is justified because according to observations W and,
particularly, C concentrations ($<$ 0.9 at. \%) are quite small in the binder phase~\cite{Linder2018}.
Therefore, the interaction between dissolved W and C in the binder 
phase can be neglected.
We calculate $ \delta{\mu}_{W(C)}$ then as the difference in total magnetic moment
between an ideal binder and a binder in which W(C) diffusion has occurred 
 
\begin{subequations}
	\begin{equation}
	\delta{\mu}_{W} = \sum_{i}^{N_{s}^{b}} \mu_{i}-\sum_i^{N_{s}^{b}} \mu_{i}^b 
\label{eq:deltaW}
	\end{equation}
	and
	\begin{equation}
	\delta{\mu}_{C} = \sum_{i}^{N_{s}^{b}} \mu_{i}-\sum_i^{N_{s}^{b}} \mu_{i}^b  
\label{eq:deltaC}
	\end{equation}
\end{subequations} 
Here, $ N_{s}^{b} $ denotes the number of binder atoms in the supercells 
shown in Fig.~\ref{fig:NiWC}, namely $N_{s}^{b}=107$
in Fig.~\ref{fig:NiWC}a) and $N_{s}^{b}=108$ in Fig.~\ref{fig:NiWC}b). 
The magnetic moments of binder atoms with ($\mu_i$) and 
without W and C diffusion ($\mu_i^b$) can be
obtained from ab-initio calculations carried out for the supercells in Fig.~\ref{fig:NiWC}.  

\subsection{\label{sec:definitions}Ab-initio Method}

Most of the calculations are performed using a method based on density functional theory (DFT)~\cite{hohenberg1964p,kohn1965self}, the exact muffin-tin orbitals 
(EMTO) method~\cite{vitos2001anisotropic,vitos2007computational}. 
In the present work, we choose the binder 85Ni15Fe as an example since
it is a promising candidate for alternative binders in cemented carbides. 
In order to simulate the binder alloy 85Ni15Fe, we use the coherent potential approximation (CPA)~\cite{vitos2001anisotropic,soven1967coherent}. CPA is a well-known approach that allows to treat random alloys within the single-site approximation~\cite{vitos2007computational}. 
In the self-consistent calculations, the one-electron equations are solved within the scalar-relativistic approximation and the soft-core scheme. \textit{s}, \textit{p}, \textit{d}, \textit{f} orbitals are included in the muffin-tin basis set. The generalized-gradient approximation (GGA) of Perdew-Burke-Ernzerhof (PBE)~\cite{perdew1996generalized} is employed as the exchange-correlation functional. The potential sphere radius of interstitial carbon is optimized to be 0.77~$w_{\mathrm{C}}^{\mathrm{0}}$, where $w_{\mathrm{C}}^{\mathrm{0}}$ 
is the atomic sphere radius of the corresponding Voronoi polyhedron around the C site. 
The coordinates of the first nearest neighbours around C in the EMTO method are set according to the optimized configuration of C-interstitial pure Ni using Vienna Ab-initio Simulation Package (VASP)~\cite{kresse1993ab,kresse1999ultrasoft}. 
The relaxation in the W-substitutional supercell is not considered here since the 
W substitution only induces slight expansion of its first nearest 
neighbours (10\% of the C-induced expansion).
Cell size tests were performed and we found that the results were fully converged for 
3$\times$3$\times$3 supercells.
K-point grids were also carefully tested to ensure convergence 
and a grid of 5$\times$5$\times$5 was used in the calculations.
In the present work, the segregation energy calculations are performed by EMTO and 
the interface calculations by VASP.
 
%

\section{\label{sec:Results}Results and discussion}

In order to calculate the weight specific magnetization using Eqs.~\ref{eq:deltaW}
and \ref{eq:deltaC}
we used the computational models in Fig.~\ref{fig:NiWC}a) and b). The optimized lattice parameters for W-substitutional and C-interstitial supercells are both 3.55 \AA. Table~\ref{table2}
lists the calculated values for the magnetic moment of W ($\mu_{W}$), the W-induced magnetic moment change
in the binder host ($\delta{\mu}_{W})$, the magnetic moment of C ($\mu_{C}$), the C-induced magnetic moment change
in the binder host ($\delta{\mu}_{C}$) and the average magnetic moment of the binder atoms ($\mu^{b}$). 
The calculated value for $\mu^{b}$ in the bulk is 0.95 $\mu_{B}$, 
which is in perfect agreement with the Slater-Pauling curve~\cite{kubler2017theory}.
\begin{table}[t]
\caption{Ab-initio calculated values of the magnetic moments of W, the W-induced magnetic moment change
in the binder host, the magnetic moment of C, the C-induced magnetic moment change
in the binder host and the average magnetic moment of the binder atoms.}
\begin{tabularx}{0.480\textwidth}{csssss} \\\toprule 
& $\mu_{W}$ & $\delta{\mu}_{W}$ & $\mu_{C}$ & $\delta{\mu}_{C}$ & $\mu^{b}$  \\[3pt] 
& $(\mu_{B})$ & $(\mu_{B})$ & $(\mu_{B})$ & $(\mu_{B})$ & $(\mu_{B})$ \\[3pt]\hline
85Ni15Fe & $\,$-0.37 & -3.98  & -0.08  & -2.96  & 0.95  \Tstrut \\\bottomrule
\label{table2}
\end{tabularx}
\end{table}
We then obtain the following expression by inserting these calculated values into Eq.~\ref{COM1}

\begin{equation}
  \begin{aligned}
 \frac{\sigma_s}{\sigma^b_s} =1  & - 4.58 \left (\frac{C_W}{1-C_W-C_C}\right) 
  \\  &   \qquad  -3.19 \left (\frac{C_C}{1-C_W-C_C}\right).
 \end{aligned}
\label{COM2}
\end{equation}
\vspace{.2cm}

\noindent We note here that the effect of W- and C-diffusion is 
to reduce the weight specific magnetic saturation. 
Moreover, the influence of W on the magnetic moment is only significant 
to the limit of its third-nearest neighbours.

The magnetic saturation of cemented carbides measured by instruments in industry 
is obtained in units of the intrinsic Co weight specific magnetization~\cite{Linder2018}. 
Here, in order to compare directly with industrial measurements, we also express the magnetic 
saturation in the same unit, and thus we multiply Eq.~\ref{COM2} by the following factor 
 
\begin{equation}
F_{\text{Co}} = \left (\frac{\mu^b}{\mu_{\text{Co}}}\right) \frac{m_{\text{Co}}}{m^{b}} C^{b}.
\label{COM3}
\end{equation}

In Eq.~\ref{COM3}, $m_{\text{Co}}$ is the atomic mass of Co, 
$\mu_{\text{Co}}=1.67 \mu_{B}$ 
is the atomic magnetic moment of Co, $C^{b}= M^{b}/M^{\text{tot}}$ and $M^{b}$ and $M^{\text{tot}}$
are the masses of the binder and sample, respectively.    

\begin{figure}[t]
	\begin{center}
		\begin{tabular}{c@{\qquad}c}
			\resizebox{0.95\columnwidth}{!}{\includegraphics[clip]{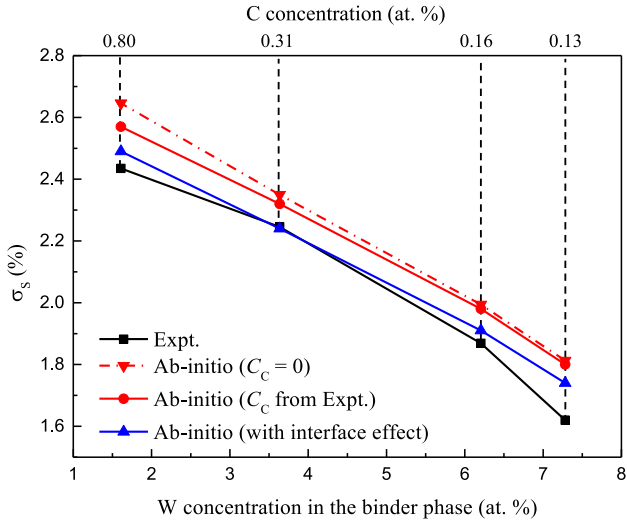}}
			\\
		\end{tabular}
\caption{\label{fig:CoM} Weight specific magnetic saturation
calculated using Eqs.~\ref{COM2} and \ref{COM3} is shown by upside down 
triangles when C content is neglected and
by circles when C concentration is taken from experimental estimations~\cite{Linder2018}. 
Triangles are used to indicate magnetic saturation including interface effects and the C content is taken from experiments.
The experimental measurements~\cite{Linder2018} are denoted by squares.}
	\end{center}
\end{figure}

Eq.~\ref{COM2} contains two variables $C_W$ and $C_C$ that are known to be inversely related. However,
the exact relationship may vary depending on many parameters, namely, grain size, diffusion rates, etc.   
To first approximation we ignore the carbon effect 
in Eq.~\ref{COM2} as it was done in Ref.~\cite{roebuck1984influence} 
and we show the magnetic saturation as a function of W concentration 
by the dash-dotted line in Fig.~\ref{fig:CoM}. 
The magnetic saturation measurements have been performed on 
NiFe-WC using a vibrating sample magnetometer by Linder \textit{et al.}~\cite{Linder2018,Hou2018} 
and are depicted by squares in Fig.~\ref{fig:CoM}. 
They estimated $C_W$ and $C_C$ from a kinetic simulation method~\cite{Hou2018,Walbruhl2017}.
We also employed their estimated $C_W$ and $C_C$ values to calculate 
the magnetic saturation and the results are presented by circles in 
Fig.~\ref{fig:CoM}. 

We first observe that the calculated weight specific magnetic saturation decreases with 
W concentration in the binder phase and that the agreement between theory and
experiments is remarkably good. Moreover,
by comparing the curves with and without C effect, it is clear that for small concentrations 
of C its influence on the magnetic saturation is negligible. 
Eq.~\ref{COM2} can then be used to predict the W concentration in the binder phase 
and hence it make it possible to predict the microstructural properties of hard metals.    

During this analysis we have neglected the effects of segregation and 
interfaces. Thus, in the following sections, 
we will investigate the contribution of these two effects separately. 

\subsection{Segregation effect on magnetic properties}

\vspace{.2cm}
In order to estimate the effect of segregation in the binder alloy we will here
develop a simple model that can be evaluated by means of our simulation method.
The coherent potential approximation used in this investigation
treats the binder alloy, 85Ni15Fe, as
an ideal effective medium. 
However, in practice, segregation of the binder elements to the dissolved W 
or C may be present. 
Such segregation effects could alter the calculated values for the 
magnetic saturation due to the different magnetic responses of W and
C to their neighbours: Ni and Fe (see the upper part of Table~\ref{table:Mag}). 
The segregation energy of the 85Ni15Fe alloy is calculated following the
work of Delczeg \textit{et al.}~\cite{delczeg2012ab} and using the 
supercells in Fig.~\ref{fig:NiWC},

\begin{multline}
E_{\text{segr}}= \dfrac{dE_{\text{1nn}}(\text{Ni}_{0.85+x}\text{Fe}_{0.15-x})}{dx}\at[\bigg]{x=0} \\
-\dfrac{dE_{\text{Bulk}}(\text{Ni}_{0.85+x}\text{Fe}_{0.15-x})}{dx}\at[\bigg]{x=0},
\label{eq:Segr}
\end{multline}
where $x$ is in this case the atomic concentration change for Ni and Fe
and it is chosen to be 0.001. The two derivatives in Eq.~\ref{eq:Segr} are calculated at constant volume. The total energy of the supercell
is denoted by $E_{1nn}(\text{Ni}_{0.85+x}\text{Fe}_{0.15-x})$ if only
the first nearest neighbours of the impurity W(C) are replaced by
$\text{Ni}_{0.85+x}\text{Fe}_{0.15-x}$, whereas if instead all sites in the supercell 
are modified, the total energy corresponds then to
$E_{\text{Bulk}}(\text{Ni}_{0.85+x}\text{Fe}_{0.15-x})$. According to this definition, 
a positive $E_{\text{segr}}$ indicates that Fe prefers to be neighbour to W(C) instead of Ni.  

The segregation energy is calculated for the ferromagnetic state and 
it is 11.1 mRy in the case of substitutional W and 12.4 mRy in the case of interstitial C. 
This indicates that Ni tends to stay far away from the dissolved W and C atoms. 
However, one should keep in mind that the segregation 
unavoidably lowers the entropy of the binder phase. 
Therefore, we also calculate the entropy of a totally disordered 85Ni15Fe-W(C) alloy,

\begin{equation}
S = n_{1n} \, c_{\text{Ni}} \ln{c_{\text{Ni}}} + n_{1n}\,  c_{\text{Fe}} \ln{c_{\text{Fe}}},
\label{eq:entropy}
\end{equation}
where $ n_{1n} $ is the number of the first nearest neighbours of W(C). $ n_{1n} $ is equal to 12 and 6 in W-substitutional and C-interstitial systems, respectively. 
$ c_{\text{Ni}} $ and $ c_{\text{Fe}} $ are the atomic concentrations of Ni and Fe, respectively. 
The entropy is calculated to be -3.21$\times$$ 10^{-2}$  mRy/K
and -1.61$\times$$ 10^{-2} $ mRy/K for W-substitutional and C-interstitial 85Ni15Fe alloys, respectively. 
The temperatures, above which the entropy of the binder phase 
dominates the total energy, are then 346 K and 770 K, respectively. 
This means that the entropy contribution is dominating at both,
 the solidification temperature and the temperature where solid state 
diffusion stops in the case of W. 
Some segregation could still be expected in the C case but since 
the amount of dissolved C is very low in the binder phase, we conclude 
that segregation does not contribute significantly to the magnetic 
saturation.

\subsection{\label{sec:interface}Interface effect on magnetic properties}
\vspace{.2cm}

\begin{figure}[t]
	\begin{center}
		\begin{tabular}{c@{\qquad}c}
			\resizebox{0.95\columnwidth}{!}{\includegraphics[clip]{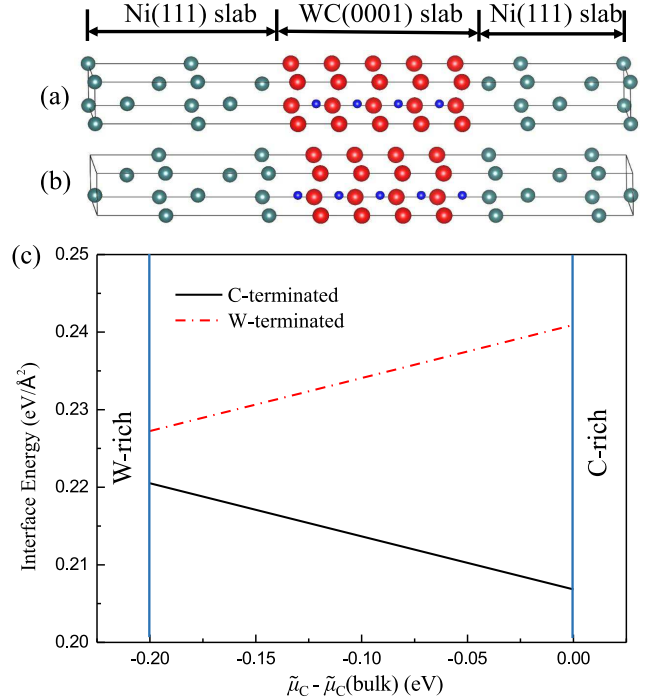}}
			\\
		\end{tabular}
		\caption{\label{fig:Ni_WC_inter} (a) W-terminated and (b) C-terminated Ni(111)/WC(0001) interfaces. The red, blue and cyan balls represent W, C and Ni atoms, respectively. (c) the calculated interface energy of W-terminated and C-terminated Ni(111)/WC(0001) interfaces as a function of the chemical potential of C, $\tilde{\mu}_{\text{C}}$.}
	\end{center}
\end{figure}

The weight fraction of the binder phase is relatively low ($\sim$5 wt. \%), which means that many atoms from the binder phase in fact sit close to the hard phase and thus might possess different magnetic state as compared to the bulk binder atoms. Here we address this question and estimate the binder/WC interface effect on the magnetic saturation. 
The magnetic behaviour of the binder elements near the interface is studied 
by constructing an interface model Ni(111)/WC(0001). 
This interface model consists of the low-index Ni(111) surface in fcc Ni 
and WC(0001) surface in hexagonal close packed (hcp) WC as 
shown in Fig.~\ref{fig:Ni_WC_inter}a). Here, the
Fe/WC interface is not considered since 
Ni is the dominant component in the binder 85Ni15Fe. 

Two Ni(111)/WC(0001) interface models with W- and C-terminated WC(0001) surfaces 
are considered and are shown in Fig.~\ref{fig:Ni_WC_inter}(a) and (b), respectively. 
The Ni/WC interfaces are modelled using the coherent interface 
approximation~\cite{li2011theoretical}, in which the averaged lattice parameter from both surfaces is used for the interface models. 
Accordingly, a = 2.70 \AA
is adopted as the new lattice parameter of the Ni/WC interface,
 while the individual c/a ratios of Ni(111) and WC(0001) surfaces are fixed. 
The mismatch between Ni(111) and WC(0001) 
surfaces is approximately 16.8\%. 
Although, this mismatch is large we expect that the magnetic moments of bulk Ni are
not severely affected as it can be seen in Table~\ref{table:Mag} ($\sim$0.62 $\mu_{B}$ in the equilibrium bulk calculation and $\sim$0.70 $\mu_{B}$ in the interface calculation).  
The interface geometry follows the stacking sequence of Ni(111) surface. 
The atoms near the interface are relaxed. 

\begin{table*}[t]
	\caption{\label{table:Mag} Magnetic moments of Ni, Fe, W, C in 
the bulk fcc model (Fig.~\ref{fig:NiWC}) 
and Ni, W, C in the interface model (Fig.\ref{fig:Ni_WC_inter}). Ni* and Fe* denote the Ni and Fe atoms far away from the impurity.}
\vspace{.1cm}	\begin{tabularx}{\textwidth}{XllXXXXXX}
		\toprule 
		\multirow{3}{*}{Model} & & \multicolumn{7}{c}{Magnetic moment ($\mu_{B}$)} \\[2pt]
		\cline{3-9}
		&\multicolumn{2}{c}{\rule[-1.5mm]{0mm}{5.5mm}}& Ni-1n & Fe-1n & Ni* & Fe* & C & W  \\
		\midrule
		\multirow{2}{*}{Bulk}& C-alloying & & 0.28 & 2.17 & 0.63 & 2.78 &  -0.08 & -- \\ 
		&W-alloying & & 0.38 & 2.47 & 0.62 & 2.78 & -- & -0.37 \\
		\midrule
		\multirow{2}{*}{Interface}& C-termination &  & 0.03 & -- & 0.70 & -- & 0.00 & -- \\ 
		& W-termination & & 0.24 & -- & 0.70 & -- & -- & 0.02\\	
		\bottomrule
	\end{tabularx}
\end{table*}  

In order to investigate the interface stability 
we calculate the interface energy $\gamma_{\text{Ni/WC}}$ following the work
in Ref.~\cite{christensen2004effects}. 
The interface energy is then calculated as 

\begin{equation}
\gamma_{\text{Ni/WC}}(\tilde{\mu}_{C})=\gamma_{\text{Ni}}+\gamma_{\text{WC}}(\tilde{\mu}_{C})-W_{\text{sep}},
\label{eq:inter}
\end{equation}

where $W_{\text{sep}}$ is the ideal work of separation of the interface,
$\gamma_{\text{Ni}}$ and $\gamma_{\text{WC}}(\tilde{\mu}_{\text{C}})$ are 
free surface energies of Ni(111) and WC(0001), respectively. The free 
surface energy of WC(0001) is a function of the chemical potential of carbon,
$\tilde{\mu}_{\text{C}}$, according to Siegel's work~\cite{siegel2002adhesion}.
Since the C composition in the binder phase is unknown, 
the exact chemical potential of C is also unknown before hand,
however we know that the C content should be between the $\eta$ and graphite phase.
Subsequently, the chemical potential of C in graphite is taken as reference 
and $\tilde{\mu}_{C} $ - $\tilde{\mu}_{C} $(bulk) is taken to be $\sim$0.2 eV as in 
the WC-Co alloys~\cite{guillermet1989thermodynamic}. 
The work of separation is defined as

\begin{equation}
W_{\text{sep}}= \frac{1}{2A}(E_{\text{Ni}}+E_{\text{WC}}-E_{\text{int}}),
\label{eq:work_sep}
\end{equation}
where $ E_{\text{Ni}}$ and $ E_{\text{WC}}$ 
are the total energies of Ni and WC slabs, respectively, $E_{\text{int}}$ 
is the the total energy of the interface Ni(111)/WC(0001)/Ni(111) and 
A is the area of the interface.  The free surface energies of Ni(111) and WC(0001)
in Eq.~\ref{eq:inter} are computed using the following expressions

\begin{equation}
\begin{aligned}
\gamma_{\text{Ni}} & = \frac{1}{2A}(E^s_{\text{Ni}}-N_{\text{Ni}} \, \tilde{\mu}_{\text{Ni}}) 
\qquad {\text{and}}\\
\gamma_{WC}(\tilde{\mu}_{C}) & =\frac{1}{2A} (E^s_{\text{WC}}-N_{W} \, \tilde{\mu}_{\text{WC}})  \\
& \qquad \qquad + \frac{1}{2A} (N_{W}-N_{C})\tilde{\mu}_{C},  
\label{eq:surf_WC}
\end{aligned}
\end{equation}
where $ N_{\text{Ni}}$ is the number of Ni atoms in Ni(111) surface model, 
$ N_{W} $ and $ N_{C} $ 
are the numbers of W and C atoms in WC(0001) surface model, respectively. 
In Eq.~\ref{eq:surf_WC} $\tilde{\mu}_{\text{Ni}}$ and
$\tilde{\mu}_{\text{WC}}$
are the chemical potentials of bulk Ni and the WC unit in bulk WC, respectively.
$E^s_{\text{Ni}}$ and $E^s_{\text{WC}}$ correspond to the total energies of Ni(111) and
WC(0001) surfaces, respectively.

The calculated interface energies for C-terminated and W-terminated Ni/WC interfaces 
are presented in Fig.~\ref{fig:Ni_WC_inter}(c) as a function of chemical
potential of carbon. One can observe that the C-terminated Ni/WC interface is 
thermodynamically more stable than the W-terminated interface over the whole 
range of carbon chemical potential considered in the present work. 
  
 \begin{figure*}[t]
	\begin{center}
		\begin{tabular}{c@{\qquad}c}
			\resizebox{0.85\textwidth}{!}{\includegraphics[clip]{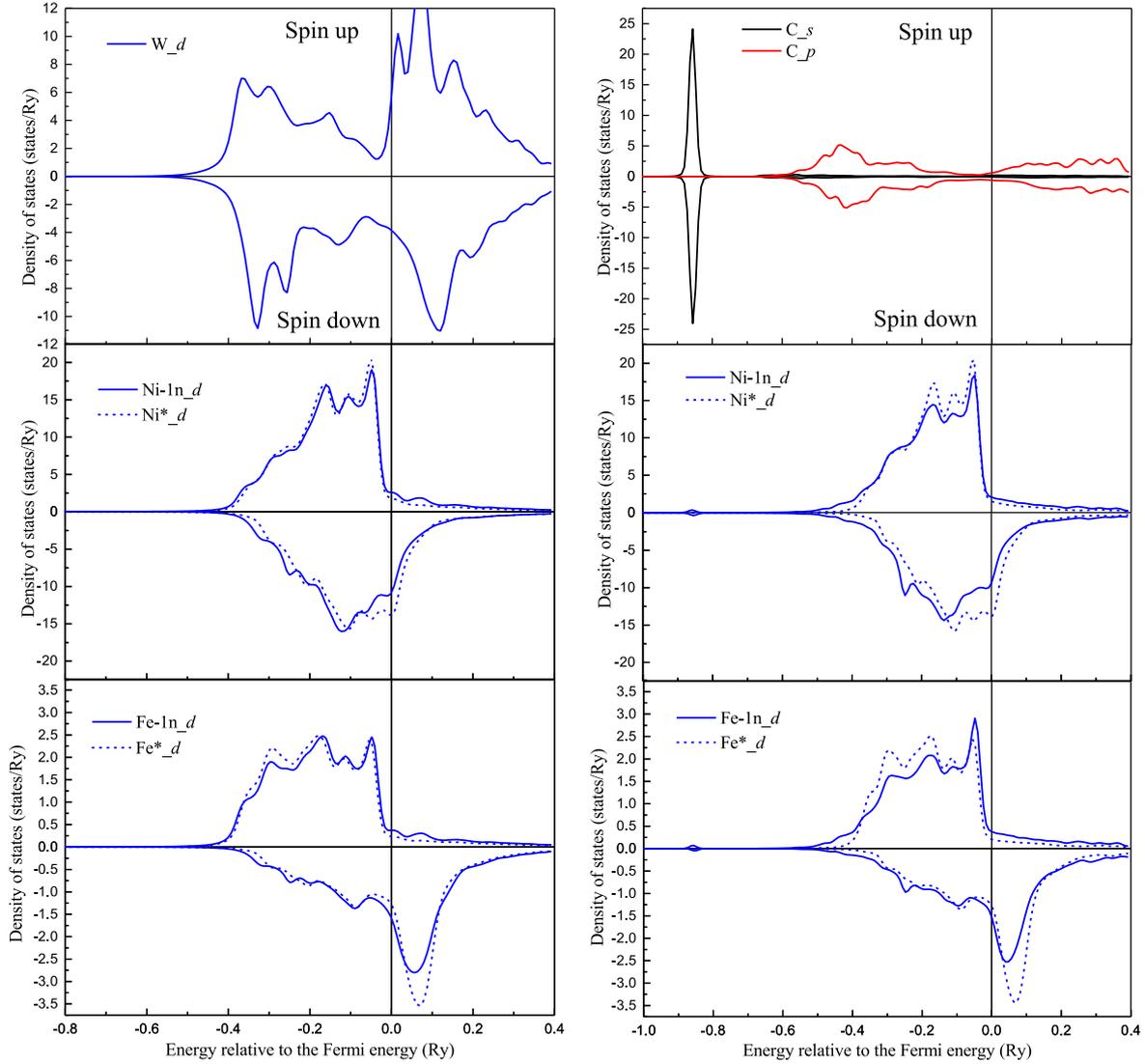}}
			\\
		\end{tabular}
		\caption{\label{fig:COM_DOS} (a) Calculated density of states for W, Ni and Fe 
in the W-substitutional bulk fcc model (Fig.~\ref{fig:NiWC} a) and  
(b) calculated density of states for C, Ni and Fe in the C-interstitial bulk fcc model 
(Fig.~\ref{fig:NiWC} b). Fe-1n and Fe$^*$ correspond to first nearest
neighbor to the impurity and an atom far from the impurity, respectively.}
	\end{center}
\end{figure*}

The effects of C and W on the magnetic moments 
of Ni around the interface can be seen in the lower part of Table~\ref{table:Mag}. 
Table~\ref{table:Mag} shows that both C and W decrease the magnetic moments 
of the Ni atoms located at the interface. The atomic magnetic moments of the interfacial Ni 
atoms are reduced to 0.03 $\mu_{B}$ and 0.24 $\mu_{B}$ for C-terminated 
and W-terminated interface models, respectively, from the value 0.70 $\mu_{B}$ when Ni is far away from the interface. 
The magnetic moments of C and W near the interface are 0 $\mu_{B}$ and 0.02 $\mu_{B}$, 
respectively. 

To estimate the upper limit of the
interface effect, here we assume that all the present interfaces in cemented 
carbides are C-terminated and that the magnetic moments of the binder elements 
near the interface are approximately zero. Moreover, a tetrahedral shape is assumed for the binder phase. The size of the binder lakes is 
set to be 0.1 $ \mu $m, which is in the same quantitative level as 
reported in Ref.~\cite{marshall2015role}. Based on these assumptions, 
the ratio of the number of surface atoms to the number of bulk atoms 
in the binder lake $n_{s}/n_{b}$ is calculated by 

\begin{equation}
\frac{n_{s}}{n_{b}} =\left (\dfrac{\sqrt{3} \, a^{2}}{\pi r^{2}} \right)
\dfrac{\dfrac{4\pi r^{3}}{3}}{\dfrac{\sqrt{2}a^{3}}{12}}
= \frac{\sqrt{2}a}{16\sqrt{3}r},
\end{equation}
where $a$ is the length of edge in the tetrahedron and 
$ r $ is taken for simplicity as the atomic radius of Ni since 
the atomic radii of Fe and Ni are quite similar. 

The weight specific magnetic saturation can be now evaluated 
including the calculated interface effect and the new curve can be seen in Fig.~\ref{fig:CoM} (with triangles).
One can see that the consideration of interface effect lowers the magnetic saturation 
and hence, the theoretically-predicted magnetic saturation becomes even
closer to the experimental measurements. Here, we also need to mention that by considering the 
random shape distribution of the binder lakes in cemented carbides, 
the assumed shape with a lower surface-area-to-volume ratio would give a 
smaller interface effect on the absolute magnetic saturation value. Furthermore, 
the W-terminated interface might also exist in the 
cemented carbides although our results show that it is thermodynamically less stable 
than the C-terminated interface.     

Interestingly, we notice from Table~\ref{table:Mag} that W and C 
show different magnetic behaviours. In the bulk computational model (Fig.~\ref{fig:NiWC}), 
C decreases the magnetic moment of Ni (Fe) more dramatically than W, while W 
itself gains more negative magnetic moment.  
Such a difference can be explained with their corresponding 
density of states (DOS) displayed in Fig.~\ref{fig:COM_DOS}. 
The \textit{d} electrons of W contributes to its negative 
magnetic moment while the electronic structures of its 
surrounding Ni and Fe atoms (Ni-1n, Fe-1n) 
are nearly unchanged compared to the Ni(Fe) atoms far away from the impurity
(Ni*, Fe*). In contrast, the spin-up and 
spin-down DOS are almost identical for C, indicating 
the approximately zero magnetic moment of C. 
The electronic structure of C-neighboured Ni and Fe atoms 
changes more due to the presence of C than that of 
those atoms far away from C. 
Similar magnetic features can also be observed in the interface model; 
the magnetic moment of the interfacial Ni atom (0.03 $\mu_{B}$) 
in the C-terminated interface is much lower than the one in 
the W-terminated interface (0.24 $\mu_{B}$). 
The different behaviours of W and C have also been observed in Co-W and Ni-P 
alloys~\cite{szpunar1998magnetism}.
C and W atoms in the Ni/WC interface show approximately 
zero magnetic moment due to the strong covalent bond between C and W in WC(0001) slabs. 

 
\section{\label{Conclu}Conclusions}
  
In the present work, we build a theory-based relationship between 
the weight specific magnetic saturation and the components of 
the binder phase for cemented carbides WC-85Ni15Fe. 
This relationship predicts the amount of W in the binder phase of 
cemented carbides, and it constitutes a non-destructive quality control. 
The calculated magnetic saturation is in excellent agreement with the experimentally measured 
values. 

The segregation of Ni (Fe) to dissolved W or C and the Ni-Fe/WC interface 
are investigated as well. The segregation study indicates that Ni tends 
to be segregated from the dissolved W or C in the binder phase. 
However, for the W- and C-dissolved binder phase, 
the entropy would dominate the total energy above 346 and 770 K, respectively. 
The segregation effect on the magnetic saturation is therefore believed not to be significant. 
The Ni(111)/WC(0001) interface is 
constructed and its stability is investigated from first principles. 
The interface effect on the magnetic saturation is estimated 
and by including this effect the magnetic saturation is reduced further, becoming even closer
to experimental values. 

Our results show that the magnetic behaviours of W and C are different 
in the binder phase. W decreases the total magnetic moment of the whole 
binder phase mainly by gaining itself negative magnetic moment, 
while C reduces the magnetic moments of its neighbours. Such distinct characteristics are explained by 
their corresponding electronic structure.  

The present theoretical method provides a quantum mechanical approach to build 
a non-destructive quality control process for cemented carbides. 
Moreover, the comparison between theory and experiment helps to gain a physical 
insight into the magnetic properties of cemented carbides. This theoretical method can also be applied to other 
cemented carbides with multi-component binders and therefore significantly aiding
the search for more complex binder phases for hard metals.  
\section*{Acknowledgments}

The present work is performed under the project COFREE (15048), 
funded by European Institute of Innovation \& Technology (EIT). 
The authors acknowledge the Ministry of Science and Technology (No.2014CB644001), the Swedish Research Council, the Swedish Foundation for Strategic Research, the Carl Tryggers Foundations, the Swedish Foundation for International Cooperation in Research and Higher Education, the Hungarian Scientific Research Fund (OTKA 128229), and the China Scholarship Council for financial supports. The computations were performed on resources provided by the Swedish National Infrastructure for Computing (SNIC) at Link\"oping.

\end{document}